\documentclass[12pt]{article}
\usepackage[margin=0.95 in]{geometry}
\usepackage{amsmath}
\usepackage{amssymb,amsfonts}
\usepackage[all]{xy}
\usepackage{graphicx}
\usepackage[utf8x]{inputenc}
\usepackage{amsmath}
\usepackage{amssymb}
\usepackage{float}
\usepackage{array}
\usepackage{tikz}
\usepackage{mathtools}
\usepackage{mathrsfs}
\usepackage{hyperref}
\usepackage{cite}
\numberwithin{equation}{section}
\numberwithin{equation}{section}
\numberwithin{table}{section}
\DeclareMathOperator{\tr}{tr}
\begin{document}

\thispagestyle{empty}

	\vspace*{3cm}
	{}
	
	\noindent
	{\LARGE \bf  Liouville Quantum Gravity}
	\vskip .4cm
	\noindent
	\linethickness{.06cm}
	\line(10,0){467}
	\vskip 1.1cm
	\noindent
	\noindent
	{\large \bf Songyuan Li$^{1}$, Nicolaos Toumbas$^{2}$, Jan Troost$^{1}$}
	\vskip 0.25cm
	{\em
		\noindent
	$^{1}$	Laboratoire de Physique
		de l'Ecole Normale Sup\'erieure \\
$ ^{}	\,\, $	CNRS, ENS, Université
		PSL, Sorbonne Universit\'e \\
	$ ^{}	\, \, $	Paris, France
		\\
	$^2$
 Department of Physics, University of Cyprus, Nicosia 1678, Cyprus
	}
	\vskip 1.2cm

	\vskip0cm
	
	\noindent {\sc Abstract: } We define a three-dimensional quantum theory of gravity as the holographic dual of the Liouville conformal field theory. The theory is  consistent and unitary by definition. The corresponding  theory of gravity with negative cosmological constant has peculiar properties. The quantum theory has  no normalizable $AdS_3$ vacuum. The model contains primary
	black holes  with zero spin. All states can be interpreted as
	% tweak
	BTZ geometries dressed with boundary gravitons. There is a unique universal interaction between these states consistent with unitarity and the conformal symmetry of the model. This theory of gravity, though conceptually isolated from other models of quantum gravity, is worth scrutinizing.
	
	\vskip 1cm

	\pagebreak
	
	\newpage
	\setcounter{tocdepth}{2}
	\tableofcontents
	
	\section{Introduction}

Consistent theories of quantum gravity are hard to come by. Quantum gravity in four dimensions (or higher) may be asymptotically safe \cite{Weinberg:1976xy}. String theory provides
us with a perturbative definition of quantum gravity in a number of backgrounds \cite{Scherk:1974ca,Yoneya:1974jg}. Moreover, string theory holography suggests a non-perturbative definition of quantum gravity in spaces with negative cosmological constant \cite{Maldacena:1997re}. Another fruitful approach is to study quantum gravity in three dimensions or lower \cite{Staruskiewicz,Deser:1983tn}. Still, the total present yield is a limited set of models of quantum gravity. We would like to extend the set.

In this paper, we study quantum gravity in three dimensions with negative cosmological constant. Firstly, we wish to remark that a large number of ultraviolet complete quantum theories of gravity in
three-dimensional anti-de Sitter space are provided by compactifying string theory on a space-time with an $AdS_3$ factor. These theories have a large set of degrees of freedom and account
microscopically for the Bekenstein-Hawking entropy of black holes \cite{Banados:1992wn,Banados:1992gq} in $AdS_3$ \cite{Strominger:1996sh}. Sufficient ingredients for this result are on the one hand that the effective central charge coincides
with the bare central charge that can be computed in gravity \cite{Brown:1986nw}, and on the other hand the Cardy formula \cite{Cardy:1986ie,Strominger:1997eq}. These string theories contain a graviton and a multitude of scalars, fermions, and higher spin fields. They permit a holographic description in terms of a two-dimensional conformal field theory dual \cite{Maldacena:1997re}.

An alternative approach to quantum gravity in three dimensions with a negative cosmological constant is to quantize the gravitational action directly \cite{Deser:1983tn,Deser:1983nh}.\footnote{It has been argued that this may be a misguided approach since it is as if one quantizes fields that describe the thermodynamics (in contrast to the statistical mechanics) of the system \cite{Martinec:1998wm}. } This track often exploits the fact that the three-dimensional gravity action can be rewritten in terms of a Chern-Simons action (with gauge algebra $so(2,2)$ or $so(3,1)$ for the case of negative
cosmological constant) \cite{Achucarro:1987vz,Witten:1988hc}. Recent attempts in this direction have concentrated on defining a conformal field theory dual with certain prescribed properties,
like an $SL(2,\mathbb{C})$ invariant vacuum and a degree of holomorphic factorization \cite{Witten:2007kt,Gaberdiel:2007ve,Maloney:2007ud}.
We note also that a careful quantization around a global $AdS_3$ vacuum was performed in \cite{Cotler:2018zff}, giving rise to a (non-modular invariant) theory of reparameterizations on a coadjoint orbit of the Virasoro group.

We rather take a different approach, {\em defining} three-dimensional quantum gravity to be the gravitational dual of Liouville conformal field theory. Manifestly, this is inspired by the identification between
the respective actions of these theories \cite{Coussaert:1995zp}. We propose their path integral measures to be identical as well.
The main advantage of this proposal is that the resulting theory is patently consistent and unitary, because the Liouville theory is. Another advantage is that the spectrum and correlation functions are known. We also stress a number of distinct properties.
Firstly, the Liouville conformal field theory does not have an $SL(2,\mathbb{C})$ invariant normalizable ground state. Thus, our quantum theory of gravity does not have a normalizable global $AdS_3$ ground state. The theory only has primary black hole states
% tweak
or BTZ geometries with zero angular momentum, which can be dressed with boundary gravitons. Relatedly, the effective central charge is one, independently of the value of the bare central charge. This has as a consequence that the Cardy formula will not reproduce the entropy of BTZ black holes.\footnote{More precisely, the continuous spectrum renders it hard to make a definite statement, let alone positive.}  We will argue that this is compatible with the topological nature of three-dimensional gravity.

Thus, the theory of gravity we define will be consistent but peculiar. As a consequence, the theory is conceptually isolated from many of the quantum theories of gravity that we admire.
Our aim in this paper is to proceed with the interpretation of this  theory of quantum gravity, and to investigate its properties. Hopefully, its original features will broaden our vision on the set of consistent theories of quantum gravity.

We remark that the idea  that pure quantum gravity with negative cosmological constant could be  equivalent to Liouville theory has a longer history.
Firstly, in \cite{Verlinde:1989ua}  the role of Liouville theory in two-dimensional quantum gravity was understood. The three-dimensional picture \cite{Coussaert:1995zp} was further developed in
\cite{Krasnov:2000ia,Krasnov:2002rn}, but different conclusions from ours were drawn. Point particle bulk states which are non-normalizable states in the boundary Liouville theory were added to the spectrum in the hope of restoring the Bekenstein-Hawking microscopic entropy counting, potentially at the cost of unitarity.\footnote{This has been a recurring theme: attempts were made to identify purported missing states, thus presumably destroying the consistency of the boundary conformal field theory.} The references
\cite{Kim:2014bga,Kim:2015bba,Kim:2015qoa} are very interesting. Especially, \cite{Kim:2015qoa}  quantizes pure gravity and identifies a projection that gives rise to the Hilbert space of Liouville theory, drawing conclusions that overlap with ours. The perspective is different, and room is left for adding spinning black holes \cite{Kim:2015bba,Kim:2015qoa}, a physical picture in terms of long strings instead of black holes \cite{Kim:2014bga,Kim:2015qoa}, et cetera. We briefly come back to the additional information that \cite{Kim:2015qoa}  provides in the conclusions. The papers
\cite{Jackson:2014nla,Turiaci:2016cvo} argue that Liouville theory arises as a universal subsector of irrational conformal field theories dual to three-dimensional theories of gravity (like $AdS_3$ string theory). While our perspective is different, the identification of gravitational dynamics with properties of Liouville conformal field theory is relevant to our proposal.\footnote{For other related work see \cite{Germani:2013sra}.}

The plan of the paper is as follows. We review the map from conformal field theory energy-momentum tensors to bulk metrics in section \ref{Tmetrics}. We map out  energy-momentum tensor
expectation values, exploiting meromorphy to a large degree. We then study the metrics we obtain in section \ref{metrics}. The analysis in these sections applies to all examples of Euclidean $AdS_3$ holography.   In particular, it does not depend on the microscopic completion of the bulk gravitational theory provided by the boundary conformal field theory. It does provide a useful backdrop for the center piece that follows. We recall
the properties of Liouville theory in section \ref{initialdictionary} and interpret them holographically. This provides us with a global picture of three-dimensional Liouville quantum gravity. We provide  entries of the holographic dictionary and discuss a  detailed map of the action as well as the state spaces of the theories. Euclidean quantum black hole mergers are described semi-classically as well as exactly. We summarize our findings and indicate open problems in section \ref{conclusions}.

\section{ Conformal Field Theory and Metrics}
\label{Tmetrics}
In the Euclidean, the bulk solution to Einstein's equation with negative cosmological constant is determined by the boundary data \cite{Witten:1998qj}. We consider various boundary set-ups, with operator insertions in different topologies. In the case of the  $AdS_3/CFT_2$  correspondence, the metric is determined by the expectation value of the energy-momentum tensor in the boundary conformal field theory (in the presence of  operator insertions). The  expectation value of the energy-momentum tensor of a two-dimensional conformal field theory is heavily constrained, and we analyze consequences of these generic properties for the bulk metric in this section. Most of these properties are implicitly or explicitly known -- we present an elementary review.

\subsection{The Energy-momentum Tensor and the Metric}
The Fefferman-Graham asymptotic expansion of the solution to Einstein gravity with  negative cosmological constant and given boundary data terminates quickly in three dimensions \cite{Banados:1998gg}. The technique used to determine it is the integration of the Chern-Simons equations of motion \cite{Banados:1998gg} with Brown-Henneaux boundary conditions \cite{Brown:1986nw}. The resulting metric is:
\begin{equation}
ds^2= l^2 d \rho^2 + 4 G l (-T  dz^2 - \bar{T} d \bar{z}^2) + (l^2 e^{2 \rho}+ 16 G^2 T\bar{T}e^{-2 \rho}) dz d \bar{z} \label{Banadosmetric} \, ,
\end{equation}
for a planar or spherical boundary parameterized by the coordinates $z$ and $\bar{z}$. The boundary data are captured by the expectation values $T=T_{zz}(z)$ and
$\bar{T}=T_{\bar z \bar z} (\bar z) $ of the energy-momentum tensor components of the boundary conformal field theory. Under conformal transformations, the boundary coordinates transform in a standard way to leading order in the radial expansion. These standard transformations are accompanied by subleading corrections as well as a change in the radial coordinate, to preserve the form of the metric (\ref{Banadosmetric})  \cite{Banados:1998gg}. Our goal in this section is to uncover generic properties of the metric (\ref{Banadosmetric}) using the properties of the energy-momentum tensor components.

\subsection{Energy-momentum Tensor Expectation Value}
The energy-momentum tensor component $T(z)$ is meromorphic. When inserted in a two-dimensional Euclidean conformal field theory correlation function, it gives rise to meromorphic functions of the
coordinate $z$ of the point of insertion.
We consider a planar boundary,
which we compactify to a two-sphere, and assume the presence of $n$ primary operator insertions. In order to determine the resulting bulk metric, we must compute the value of the energy-momentum tensor when inserted in the corresponding $n$-point function of the conformal field theory. The value of the energy-momentum tensor component $T$ (and likewise of $\bar T$) is considerably constrained by the property of meromorphy. Its double and single poles are known, given the conformal weight of the primary operator insertions. It is also restricted to behave as $z^{-4}$ at most for large $z$. This is for the point at infinity to be regular. Let us enumerate results for a low number of insertions $n$.

Firstly, the one-point function is zero because of conformal invariance. For the two-point function, with insertions at $z_1$ and $z_2$, we expect to be able to expand the energy-momentum tensor  expectation value in terms of a holomorphic function, and at most double pole singularities at $z_1$ and $z_2$. Moreover, the residues of the double poles are prescribed. Imposing that the energy-momentum tensor behaves as $z^{-4}$ for large $z$ fixes its value uniquely to:
\begin{equation}
\frac{\langle T(z) O_1 O_2 \rangle}{\langle O_1 O_2 \rangle} \equiv T^{(2)}(z) = \frac{h}{(z-z_1)^2} + \frac{h}{(z-z_2)^2} - \frac{2h}{(z-z_1)(z-z_2)} \, .
\end{equation}
We had to set $h_1=h_2$ to obtain a non-zero two-point function. It is easy to check that also the subleading single pole behaviour is as expected for a conformal primary.
There is an interesting special case, when
one of the insertions is at infinity, and the other insertion is at zero:\footnote{Regularity at infinity is now violated because of the insertion.}
\begin{equation}
T^{0,\infty}(z) = \frac{h}{z^2} \,  . \label{planarfixed}
\end{equation}
Next, we record the expectation value of the energy-momentum tensor component $T(z)$ in the presence of three primary conformal field theory operators with
conformal dimensions $h_i$:
\begin{equation}
T^{(3)}(z) = h_1( \frac{1}{z-z_1} - \frac{1}{z-z_2})(  \frac{1}{z-z_1} - \frac{1}{z-z_3}) + \mbox{cyclic} \, .
\end{equation}
Higher point functions are also heavily constrained.

\subsection{Energy-momentum Tensor on the Cylinder}
In the following we  also make use of a boundary manifold which is a cylinder. In two dimensions, it can be obtained by
the conformal transformation $z=e^{-iw}$ from the plane. It is important that the conformal transformation maps the points $z=0$ and $z=\infty$ to the open ends
of the infinite cylinder. The  expectation value of the energy-momentum tensor on the cylinder is determined from the value on the plane through the quasi-primary
transformation property of the energy-momentum tensor:
\begin{equation}
(\partial_z w)^2 T'(w) = T(z) - \frac{c}{12} \{w,z\} \, ,
\end{equation}
where the Schwarzian $\{ \cdot, \cdot \} $ for the map $z=e^{-iw}$ evaluates to:
\begin{eqnarray}
\{ w,z \} &=& \frac{1}{2 z^2} \, .
\end{eqnarray}
Thus, the energy-momentum tensor component $T_c$ on the cylinder equals
\begin{equation}
T_{c}(w)  = -z^2 T(z) + \frac{c}{24} \, .
\end{equation}
Using this map, we straightforwardly compute the expectation values of the energy-momentum tensor on the cylinder for the configurations discussed previously:
\begin{eqnarray}
T_{c}^{0,\infty} (w) &=& -h+\frac{c}{24} \, ,  \label{cylinderconstant} \\
T_{c}^{(2)} (w)
&=& \frac{h}{4} \frac{ \sin^2 \frac{w_1-w_2}{2} }{ \sin^2 \frac{w - w_1}{2}\sin^2 \frac{w-w_2}{2}} +\frac{c}{24} \, ,  \label{cylindertwo} \\
T^{(3)}_{c}(w)
&=&  h_1 \frac{\sin \frac{w_2-w_1}{2}}{2 \sin \frac{w-w_1}{2} \sin  \frac{w-w_2}{2}}   \frac{\sin \frac{w_3-w_1}{2}}{2 \sin \frac{w-w_1}{2} \sin \frac{w-w_3}{2}} + \mbox{cyclic} + \frac{c}{24} \, .
\label{cylinderthree}
\end{eqnarray}
Thus, we have computed examples of energy-momentum tensor expectation values that universally arise in two-dimensional conformal field theories. We commence the study of the corresponding metrics in the next section, and continue the analysis in Appendix \ref{LorentzianMetrics}.

\section{Universal Metrics}
\label{BHmetrics}
\label{metrics}
In the previous section, we computed universal expectation values of the energy-momentum tensor. Through the elliptic boundary value problem of three-dimensional Euclidean Einstein gravity with negative cosmological constant \cite{Witten:1998qj}, these translate into universal metrics \cite{Banados:1998gg} which we study in the present section.

\subsection{The BTZ Black Hole Metric}
We recorded the constant energy-momentum tensor expectation value  $T_c^{0,\infty}$ (\ref{cylinderconstant}) on the cylinder corresponding to a primary state of conformal dimension $h$, propagating from the infinite past to the infinite future. The resulting metric is straightforwardly determined. If we take equal left and right conformal dimensions, then the metric can be transformed to the Euclidean BTZ black hole metric with zero spin:
\begin{equation}
ds^2_{BTZ} = l^2 (\frac{r^2}{l^2} - 8 GM) dt^2 + (\frac{r^2}{l^2}-8 GM)^{-1} dr^2 + r^2 d \phi^2 \, . \label{EuclideanBTZ}
\end{equation}
The equivalence is demonstrated through the identifications and coordinate transformations \cite{Banados:1998gg}:
\begin{eqnarray}
T_{c}^{0,\infty} = \bar{T}_{c}^{0,\infty} &=&- \frac{M l}{2}  \nonumber \\
w &=& \phi + i t
\nonumber \\
r_+^2 &=& 8 G M l^2 = 4 l^2  e^{2 \rho_0}
\nonumber \\
r^2 &=& 8 GM l^2 \cosh^2 (\rho-\rho_0) \, .
\end{eqnarray}
In the classical theory we need to suppose that the conformal dimension of the operator on the plane satisfies the bound $h \ge c/24$.
The Euclidean metric (\ref{EuclideanBTZ}) can be analytically continued to the Lorentzian BTZ black hole.
This whole analysis is easily generalized to the case of a BTZ black hole with non-zero angular momentum  \cite{Banados:1998gg}.

\subsubsection*{The Horizon}

We can also study the metric if we start out with a planar boundary and the energy-momentum tensor $T^{0,\infty}$ (\ref{planarfixed}). A useful and generic tool of analysis is to determine the location of the event horizon, where the metric (\ref{Banadosmetric})
degenerates. It satisfies the equation:
\begin{equation}
e^{2 \rho_{hor}} = 4 G l^{-1} (T \bar{T})^{\frac{1}{2}} \, . \label{horizonequation}
\end{equation}
Thus, for instance, for the planar energy-momentum tensor $T^{0,\infty}$, the horizon radius is
\begin{equation}
e^{2 \rho_{hor}} = 4 G l^{-1} h \frac{1}{|z|^2} \, ,
\end{equation}
as a function of the planar coordinate $z$. The first thing to note is that the horizon now reaches radial infinity (at the boundary insertion).
If we put an infrared cut-off $R$ on the coordinate radius $e^{ \rho}$ of the bulk $AdS_3$ space, then we cut out a little hole from the surface of the event horizon, parameterized by the angle of $z$, where $R^2=4 Gl^{-1} h |z|^{-2}$.   There is an analogous singularity at $z=\infty$. Of course, the excisions correspond to the incoming and outgoing state that is present in the cylindrical set-up. In other words, the horizon surface is a sphere with two marked points, where the horizon reaches the boundary, or a sphere with two small disks cut out, when we introduce an infrared cut-off in the bulk. We recuperate the standard cylinder topology of the BTZ horizon in this manner.

\subsection{Properties of the Metrics}

It should be clear that this reasoning is very generic. The singular behaviour of the operator product expansion of the energy-momentum tensor with a (conformal primary)  operator insertion will guarantee that near a local insertion with sufficiently high conformal dimension, the metric can be locally transformed (through the exponential map) into a BTZ black hole metric with mass
$Ml = 2h-c/12$ (where  for simplicity only we assume zero spin). This is independent of the number and the precise nature of the insertions.
Thus, for example, in the cases of the energy-momentum tensors $T^{(2)}$ and $T^{(3)}$, we can state that locally, near one of the insertions, the metric looks like a BTZ black hole metric
with (left) conformal dimension $h_i$. The horizon looks like a sphere with two or three peaks, reaching the $AdS_3$ boundary.

On the cylinder, we locally find the same picture.  Care should be taken because of the fact that the unit operator gives rise to a constant energy-momentum tensor $T_c$ at infinity. The non-zero horizon radius for the cylindrical horizon should not be confused with a black hole insertion. Otherwise, the cylindrical picture is the same as the planar picture -- the reasoning can be applied locally.
Note that the singularities in the metric only depend on the planar or cylindrical coordinates on a radial slice. Thus, they propagate radially inward towards the center on straight lines.

Let us make a few more generic remarks on the universal metrics. We note that on the horizon surface defined by equation (\ref{horizonequation}), the determinant of the metric has a double zero. Thus, when we cross the surface, the nature of the metric remains the same, in the sense that we are still outside the horizon. One can see this in more detail in e.g. the BTZ metric, in which case the radial coordinate $\rho$ touches the horizon at $\rho_0$, and then returns to infinity.

In appendix \ref{LorentzianMetrics}, we start the exploration of the metrics associated to the energy-momentum tensors $T^{(2)}$ and $T^{(3)}$ after analytic continuation to the Lorentzian
cylinder.
We believe these Euclidean and Lorentzian metrics are interesting, and since they are universal by meromorphy, they deserve further analysis. As background for the following section, it is more useful to us to turn to a different characterization of the same class of metrics, in terms of monodromies.

\subsection{On Monodromies}
\label{connections}
\label{monodromies}
It is known how the above picture is  reflected in the Chern-Simons formulation of three-dimensional gravity. The Banados metric (\ref{Banadosmetric}) is associated to the (left) Chern-Simons connection \cite{Banados:1998gg}:
\begin{eqnarray}
A &=& \left( \begin{array}{cc} - \frac{1}{2} d \rho & -i e^\rho dw \\
- i \frac{4GT}{l} e^{-\rho} dw & \frac{1}{2} d \rho \end{array} \right) \, .
\end{eqnarray}
This connection is the left combination of dreibein and spin connection (and there is an analogous right combination).
The presence of BTZ sources in the metric is tracked by the monodromies of the connection around the singularity \cite{Banados:1992gq}. The simplest example is again the standard BTZ black hole, where
the left connection reads
\begin{eqnarray}
A &=& \left( \begin{array}{cc} - \frac{1}{2} d \rho & -i e^\rho dw \\
2 i GM e^{-\rho} dw & \frac{1}{2} d \rho \end{array} \right) \, .
\end{eqnarray}
We integrate it over the angular part of $w$, at the horizon radius $e^{\rho_0}=\sqrt{2 GM}$. The matrix is then constant and gives rise to a monodromy trace
\begin{equation}
\tr P \exp  \int A  =2 \cosh 2 \pi \sqrt{2 GM} \, .
\end{equation}
Our next example is slightly more involved. We compute the monodromy around a Euclidean insertion on the plane. The left-connection equals
\begin{eqnarray}
A &=& \left( \begin{array}{cc} - \frac{1}{2} d \rho & -i e^\rho dz \\
- i \frac{4G}{l} \frac{h}{z^2} e^{-\rho} dz & \frac{1}{2} d \rho \end{array} \right)  \, .
\end{eqnarray}
At fixed $e^{2 \rho_{hor}} = \frac{4G}{l} h/|z|^2$, the connection reduces to
\begin{eqnarray}
A
&=& \left ( \begin{array}{cc} 0 & \sqrt{\frac{4G}{l} h} e^{i \theta} d \theta \\
\sqrt{\frac{4 G}{l} h} e^{-i \theta} d \theta & 0 \end{array} \right)
	\, .
\end{eqnarray}
We can gauge transform the connection to a constant matrix, and compute the resulting trace:
\begin{equation}
\tr P \exp \int A = 2 \cosh 2 \pi \sqrt{\frac{4 G}{l}  (h-c/24)} \, .\label{insertionmonodromy}
\end{equation}
Thus we recognize Euclidean vertex operator insertions as BTZ geometries,  locally.
More generally,  near a Euclidean insertion that gives rise to a double pole in the energy-momentum tensor component $T(z)$, the monodromy calculation and result reduces to the hyperbolic
monodromy  (\ref{insertionmonodromy}).

\section{Liouville Quantum Gravity}
\label{initialdictionary}
In the previous sections, we discussed a few universal properties of the $AdS_3/CFT_2$ holographic duality. In this section, we commit to a very particular dual conformal field theory, namely Liouville theory. The boundary conformal field theory is then well studied. We take the holographic dictionary to {\em define} the bulk three-dimensional theory of quantum gravity, and call the resulting theory Liouville quantum gravity. In this section, we first recall  properties of the unitary Liouville conformal field theory.  We then provide a holographic gravitational interpretation of these basic properties, and critically discuss the ensuing theory of quantum gravity.

\subsection{Properties of Liouville Theory}
Liouville theory is a consistent unitary conformal field theory in two dimensions. See e.g.  \cite{Seiberg:1990eb,Teschner:2001rv} for an older and a more recent review.
The spectrum and the three-point functions entirely determine the theory. The consistency of the
theory hinges on crossing symmetry of the four-point function on the sphere and the covariant modularity of the torus one-point function. Both have been understood \cite{Ponsot:2000mt,Teschner:2003at}.
 Therefore, if we define three-dimensional quantum gravity as the dual of
Liouville theory, it will be manifestly consistent as well as unitary. The question arises which properties the resulting theory of quantum gravity exhibits when expressed as a three-dimensional bulk theory through holographic duality.
To answer the question, we firstly review various properties of Liouville theory and then translate them using a holographic dictionary.

The Liouville action $S_L$  is
\begin{equation}
S_{L} = \frac{1}{4 \pi} \int d^2 z \left( \partial \phi \bar{\partial} \phi + \mu e^{2 b \phi} + Q R^{(2)} \phi  \right) \, , \label{Liouvilleaction}
\end{equation}
where the parameter $b$ is related to the bare central charge $c$ of the theory by
\begin{equation}
c  = 1 + 6 Q^2 = 1 + (b+b^{-1})^2 \, ,
\end{equation}
which ensures  that the exponential interaction term in the action is marginal.
The conformal dimensions of the Liouville vertex operator $e^{2 \alpha \phi}$  is given by the formula
\begin{equation}
h = \alpha (Q-\alpha) \, ,
\end{equation}
and the spectrum of momenta of the unitary Liouville theory is determined by the momenta $\alpha \in Q/2 + i \mathbb{R}^+$. Thus, the continuous spectrum has a lower limit   conformal dimension $h_{min}=Q^2/4=(c-1)/24$. The effective central charge relevant to the Cardy formula is then
\begin{equation}
c_{eff} = c-24\, h_{min}=1 \, .
\end{equation}
The effective central charge reflects the fact that we have a single boson, and this is visible in the ultraviolet growth of the number of states as a function of their conformal dimension \cite{Cardy:1986ie}.

The Liouville action (\ref{Liouvilleaction}) in the presence of operator insertions needs to be regularized. In the semi-classical theory, the semi-classical Liouville field $\varphi=2b \phi$ satisfies the boundary conditions
\begin{eqnarray}
\varphi &=& -2 \eta_i \log |z-z_i|^2 + O(1) \quad \mbox{at} \quad z \rightarrow z_i
\nonumber \\
\varphi&=& -2 \log |z|^2 + O(1) \quad \mbox{at} \quad z \rightarrow \infty \, ,
\end{eqnarray}
in the presence of operator insertions with $\alpha_i \approx \eta_i/b$ and a linear dilaton charge at infinity. Following e.g. \cite{Zamolodchikov:1995aa}, we cut out small disks $D_i$ of radius $\epsilon_i$ around each insertion, as well as around infinity, and define the
regularized Liouville action:
\begin{equation}
S_{L}^{reg} = S_L +  \varphi_{\infty}  + 2 \log r_{\infty} - \sum_i (\eta_i \varphi_i + 2 \eta_i^2 \log \epsilon_i) \, , \label{regLiouvilleaction}
\end{equation}
where we defined the boundary contributions
\begin{equation}
\varphi_\infty =  \frac{1}{2 \pi r_{\infty}} \int_{\partial D_\infty} \varphi dl \, ,
\end{equation}
and
\begin{equation}
\varphi_i=  \frac{1}{2 \pi \epsilon_i} \int_{\partial D_i} \varphi dl \, ,
\end{equation}
which cancel the divergences and insure that the boundary conditions are consistent.

The Liouville three-point functions can be guessed  by computing their  pole structure and their semi-classical behaviour \cite{Dorn:1994xn,Zamolodchikov:1995aa}. The three-point functions
can also be derived
by requiring that  null vectors in  degenerate representations (reached by analytic continuation from the unitary spectrum) decouple \cite{Teschner:1995yf}. Furthermore, the three-point functions
 can  be numerically bootstrapped  \cite{Collier:2017shs}. Explicitly, they are  given by:
\begin{equation}
C(\alpha_1,\alpha_2,\alpha_3) = \frac{Y_0 Y(2 \alpha_1)Y(2 \alpha_2)Y(2 \alpha_3)}{Y(\alpha_1+\alpha_2+\alpha_3 -Q)Y(\alpha_1+\alpha_2-\alpha_3)Y(\alpha_2+\alpha_3-\alpha_1)Y(\alpha_3+\alpha_1-\alpha_2)}
\end{equation}
where the definition of the special function $Y$ is
\begin{equation}
\log Y (x) = \int_0^\infty  \frac{dt}{t} \left( (\frac{Q}{2}-x)^2 e^{-t} -  \frac{\sinh^2 (\frac{Q}{2}-x) \frac{t}{2}}{\sinh \frac{bt}{2} \sinh \frac{t}{2b}} \right) \, ,
\end{equation}
and we need:
\begin{equation}
Y_0 = \partial_x Y(x)_{|x=0}\, .
\end{equation}
%defined by
%\begin{equation}
%\log \Gamma_2(x | \omega_1,\omega_2) =  (\partial_s \sum_{n_1,n_2=0}^\infty (x+ n_1 \omega_1 + n_2 \omega_2)^{-s})_{s=0} \, .
%\end{equation}
The reflection amplitude codes the fact that an operator in Liouville theory at momentum $\alpha = Q/2 + i p$, where $p$ is a positive real number, creates the same state as the operator at momentum $Q/2-ip$. In the mini-superspace
approximation to Liouville theory, a normalizable state consists of a wave bouncing off an exponential potential, and returning with equal and opposite momentum. We have an equation of the
type
\begin{equation}
|-p\rangle = R(-p) | p \rangle \, , \label{reflection}
\end{equation}
where the reflection amplitude $R$ is related to the three-point function by taking a limit in one of the momenta. See e.g. \cite{Teschner:2001rv}  for the detailed formulas.
These data fix the Liouville conformal field theory uniquely, and it can be proven to satisfy crossing symmetry  as well as consistency of the torus one-point function. Thus, we have a two-dimensional conformal field theory that is consistent and unitary, and that to a large extent is unique, as seen both from the null vector decoupling argument \cite{Teschner:1995yf}, and from the numerical bootstrap approach \cite{Collier:2017shs}.

\subsection{The Holographic Interpretation}
In this subsection, we holographically interpret  properties of Liouville theory. We first discuss the big picture of  three-dimensional Liouville gravity, and  subsequent subsections fill in more details.

\subsubsection{The Big Picture}
The Liouville conformal field theory has peculiar properties that translate into unexpected properties of its proposed three-dimensional quantum gravity dual. Firstly, there is no normalizable $SL(2,\mathbb{C})$ invariant vacuum state in the Euclidean conformal field theory. This translates holographically into the absence of an $so(3,1)$ invariant quantum gravity ground state in the  theory. In other words, the path integral of our quantum theory of gravity is such that the   Euclidean $AdS_3$ space-time is not normalizable.
% tweak
 The spectrum of the theory consists of primaries
with conformal dimensions which form a continuous spectrum starting at a gap $(c-1)/24$. These primaries create hyperbolic monodromies at their insertion, and translate  holographically
to BTZ space-times. Moreover, the descendants, generated by the action of the Fourier modes of the boundary energy-momentum tensor, can be interpreted as boundary gravitons localized at the boundary of the three-dimensional space-time.

Thus the spectrum of the boundary conformal field theory  has the following bulk interpretation. The only excitations that we have in our three-dimensional theory of quantum gravity are  primary BTZ black holes, dressed with boundary gravitons. The theory has no propagating gravitons or other particles.
A naked black hole (i.e. with no boundary gravitons) corresponds to a single quantum state. It has zero entropy.
The boundary gravitons give entropy to a state at given high mass $Ml= L_0+\bar{L}_0$ as dictated by the Cardy formula, at effective central charge $c_{eff}=1$. \footnote{ In no sense do these degrees of freedom account fully for the traditional semi-classical Bekenstein-Hawking entropy of the three-dimensional black holes.}\footnote{We have divided by a space-time volume factor due to the continuous spectrum in making these statements. }
The theory we define only has primary states of spin zero. Thus, there are no primary spinning black holes.

In our quantum theory of gravity, it is hard to detect the horizon of a black hole. That is because there are no propagating degrees of freedom. The only degrees of freedom are the primary states corresponding to the black holes themselves, as well as the boundary degrees of freedom that live at spatial infinity. Thus, it is
% tweak
impossible to derive the traditional
% tweak
macroscopic picture of the thermodynamic properties of the BTZ black holes  \cite{Hawking:1974sw,Carlip:1994gc} within our limited theory. This may be just as well, since we have just established that the
% tweak
microscopic entropy of the black holes we consider is zero. On the other hand,
coupling to other degrees of freedom would re-establish a traditional thermodynamic picture, but it is challenging to do this consistently with the postulates of holographic duality and quantum gravity.\footnote{Relatedly, in our theory,
since there is no way we can detect
the horizon of a black hole, there is little reason to call the black hole geometry black. We could therefore always refer to the BTZ black hole as the BTZ geometry.
We will refrain from this heavy change of nomenclature, and keep referring to the BTZ geometry as being a black hole. The reader should  keep the essential conceptual caveat in mind.}\footnote{At least one way
to consistently couple matter is provided by $AdS_3$ string theory. The resulting quantum theory of gravity has a multitude of extra degrees of freedom -- enough to measure as well as account for the black hole entropy. }\footnote{Note that in string theory too, it is impossible to implement the standard low energy operation of adding a single scalar field. Our theory of quantum gravity shares this feature of being rigid.}
% tweak

The Liouville theory has one parameter which is the bare central charge $c$. A perturbation expansion exists in one over the bare central charge $c$, the $1/c$ expansion. In the dual gravitational theory, the expansion is in $G_N/l$ where $G_N$ is the three-dimensional Newton constant and $l$ is the radius of curvature of the locally $AdS_3$ space-time.

There is a detail that we skipped over in our explanation. The $AdS_3$ space-time corresponds to a state with (planar) conformal dimension $h$ equal to zero both classically and quantum mechanically.
It therefore has weight $-c/24$ with respect to the cylindrical Hamiltonian. The minimal conformal dimension
in the spectrum of Liouville theory is $Q^2/4=(c-1)/24$. In the cylinder frame, this becomes a minimal mass of  $-1/24$. This is negative. Starting at (planar) conformal dimension
$(c-1)/24$, we  already identify the resulting states with (quantum) BTZ geometries because they have hyperbolic monodromy quantum mechanically. Note that this quantum correction to the classical picture is a one-loop  effect. We will comment further on one-loop effects in subsection \ref{oneloop}.

\subsubsection{The Action}

\label{actiondetails}
Since  the holographic relation between Liouville theory and the bulk gravitational theory is crucial to us, we carefully review the map between the boundary and the bulk action  \cite{Coussaert:1995zp} in appendix
\ref{action}. We do modify the discussion of the zero modes of the bulk theory.  In particular, we gauge a  symmetry in order to guarantee that the left and right zero modes of the reduced, chiral Wess-Zumino-Witten models on the boundary are glued together, such that we reproduce the diagonal spectrum of the boundary Liouville theory. The details of the procedure are in appendix \ref{action}.

\subsubsection{The Reflection Amplitude}

Each Liouville state has a gravitational dual state. The Liouville state satisfies a reflection property (\ref{reflection}). We can ask whether there is a bulk analogue of the reflection property.
Firstly, we should note that the bulk geometry depends only on the boundary energy-momentum tensor, which is invariant under reflection. Thus, the bulk geometry is invariant
under reflection.  Thus, there is a pressing question as to how evaluating the bulk action can lead to a different result for on the one hand momentum $p$ and on the other hand momentum $-p$.

As demonstrated in appendix \ref{action}, the bulk action equals the boundary Liouville action. A subtlety arises when we allow primary operator insertions. This guarantees that the bulk and boundary action have additional divergences. Thus, we must regularize the bulk (and boundary) action in the presence of operator insertions. It would be interesting to analyze bulk regularization in the presence of  operator insertions in generality. We concentrate on the following phenomenon.

When we introduce the microscopic picture for the boundary energy-momentum tensor provided by Liouville theory, there are natural candidate primary operator insertions, defined in terms of the Liouville field.
In semi-classical Liouville theory, in order to evaluate the action, we regularize it. The regularization depends on the Louville field, which in turn does depend on the choice of the sign of the momentum. (See e.g. equation (\ref{regLiouvilleaction}).) Thus, the semi-classical (regularized Liouville, bulk, and boundary) action  changes under a change of momentum. From the bulk perspective, the regularization is in terms of a field which is a non-local functional of the energy-momentum tensor. The Liouville regulator does have neat properties. It guarantees that independently of the choice of momentum (which is, roughly, a square root of the conformal dimension), the semi-classical actions will be related by a phase factor. Thus, they can be associated to the same (albeit reflected) state (as in equation (\ref{reflection}).\footnote{See also \cite{Krasnov:2000ia} for a detailed defense of the statement that natural regulators for the bulk gravity action  lead to equality with the boundary Liouville action, albeit in the differing context of non-normalizable point particle excitations.}

The semi-classical actions for the three-point correlator (see also subsection \ref{blackholecorrelators}), for instance, are related by a factor which is the semi-classical limit $R_{s.c.}$ of the reflection amplitude \cite{Zamolodchikov:1995aa}:
\begin{equation}
e^{-\frac{1}{b^2} S_{cl}(p_1,p_2,p_3)} \approx R_{s.c.}(-p_1)  e^{-\frac{1}{b^2} S_{cl}(-p_1,p_2,p_3)} \, .
\end{equation}
Finally, quantum Liouville theory and the quantum reflection amplitude $R$ extend
the semi-classical discussion to a theory that is exact in $G_N/l$.

\subsubsection{The States}

The Liouville partition function is
\begin{equation}
Z = V \int \frac{dp}{2 \pi} \frac{e^{- \pi \tau_2 p^2 }  }{|\eta(q)|^2} = \frac{V}{2 \pi \sqrt{\tau_2} |\eta(q)|^2}\, ,
\end{equation}
where $q=e^{2 \pi i \tau}$, $\tau$ is the modulus of the rectangular torus and $V$ is an infinite  volume factor due to the Liouville zero mode.
Conformal invariance implies that the partition function will depend on the product $LT$ only where $L$ is the size of the boundary circle and $T$ is the temperature. We can express
this product in terms of the parameter $\tau_2$. The partition function coincides with the partition function of one free boson. This matches the fact that the effective number of boundary degrees of freedom is of finite central charge, equal to one.
The partition function is the simplest holomorphically factorized partition function which satisfies the requirement of reproducing the spectrum of boundary gravitons as well as matching the bare Brown-Henneaux central charge. The continuous spectrum of momenta matches geometries in the bulk space-time which have singularities cloaked by horizons.

With a finite number of effective degrees of freedom, we do not
expect a phase transition. The partition function is a continuous function of the temperature (times the volume) $\tau_2$. In this model there is no Hawking-Page transition. In fact, the partition function is independent of the bare central charge, and there is no semi-classical limit, nor a summed approximation over saddles.

\subsection{Black Hole Correlators}
\label{blackholecorrelators}
\label{mergers}
\label{threepoint}
We described how  Liouville quantum gravity only has black hole states dressed with boundary gravitons. Moreover, the theory enjoys the full conformal invariance of a two-dimensional conformal field theory. Beyond the known spectrum of conformal dimensions,  a natural observable to compute is therefore the correlation function of three operators that create (primary) black holes states. The correlators of descendent black holes, namely black holes dressed with boundary gravitons, are generated from these by the symmetries of the theory. If we consider the correlation function of more than three black holes, those can be reconstructed by factorization.

The three-point function respects global conformal invariance and is therefore proportional to
\begin{equation}
\langle O_1 O_2 O_3 \rangle = \frac{C_{123}}{|z_1-z_2|^{2h_1+2h_2-2h_3} |z_2-z_3|^{2h_2+2h_3-2h_1} |z_3-z_1|^{2h_3+2h_1-2h_2} } \, .
\end{equation}
In Liouville quantum gravity, the proportionality constant $C_{123}$ is given by the Liouville three-point function.
 The semi-classical three-point function is again determined by the regularized Liouville action, which we interpreted as a regularziation of the bulk action in the presence of  operator insertions. The  $1/c=2 G_N/(3l)$ corrections arise from integrating over the Liouville field configurations, which we consider to be the microscopic degrees of freedom of the theory. We  study these statements in more detail below.

\subsubsection{The Semi-Classical Liouville Three-point Function}
In this subsection, we  remind the reader of the calculation of the semi-classical approximation to the Louville three-point function \cite{Zamolodchikov:1995aa}. The boundary conditions on the semi-classical Liouville field $\varphi = 2 b \phi $ are
\begin{eqnarray}
\varphi &=& -2 \eta_i \log |z-z_i|^2 + O(1) \quad \mbox{at} \quad z \rightarrow z_i
\nonumber \\
\varphi&=& -2 \log |z|^2 + O(1) \quad \mbox{at} \quad z \rightarrow \infty \, ,
\end{eqnarray}
while the action is regularized as before.
%\begin{equation}
%\nonumber \\
%S_{L}^{s.c.}(1,2,3) = \dots
%\end{equation}
In the semiclassical limit where $b\to0$, the three point function in the Liouville theory with operator insertions $\exp(\eta_i\varphi(z_i))$ is given by $\exp(-\frac{1}{b^2}S_{cl})$
where $S_{cl}$  is the classical Liouville action evaluated on  the classical solution to the Liouville equation. This classical solution is given by \cite{Zamolodchikov:1995aa}
\begin{equation}
\varphi_{\eta_1,\eta_2,\eta_3}(z|z_1,z_2,z_3)=-2\log[a_1\psi_1(z)\psi_1(\bar{z})+a_2\psi_2(z)\psi_2(\bar{z})],\label{eq:classicalsolution}
\end{equation}
where
\begin{eqnarray}
\psi_1 &=& (z-z_1)^{\eta_1}(z-z_2)^{1-\eta_1-\eta_3}(z-z_3)^{\eta_3}\times\nonumber\\
& &{}_2F_1(\eta_1+\eta_3-\eta_2,\eta_1+\eta_2+\eta_3-1,2\eta_1,x)\label{eq:psi1}\\
\psi_2 &=& (z-z_1)^{1-\eta_1}(z-z_2)^{\eta_1+\eta_3-1}(z-z_3)^{1-\eta_3}\times\nonumber\\
& &{}_2F_1(1+\eta_2-\eta_1-\eta_3,2-\eta_1-\eta_2-\eta_3,2-2\eta_1,x) \, ,
\end{eqnarray}
with the variable $x$ defined as
\begin{equation}
x=\frac{(z-z_1)z_{32}}{(z-z_2)z_{31}}.\label{eq:x}
\end{equation}
The constants $a_1$ and $a_2$ are
\begin{eqnarray}
a_1^2 &=& \frac{\pi\mu b^2}{|z_{13}|^{4\eta_3+4\eta_1-2}|z_{12}|^{2-4\eta_2}|z_{23}|^{2-4\eta_1}}\times\nonumber\\
& &\frac{\gamma(\eta_1+\eta_2+\eta_3-1)\gamma(\eta_1+\eta_3-\eta_2)\gamma(\eta_1+\eta_2-\eta_3)}
{\gamma^2(2\eta_1)\gamma(\eta_2+\eta_3-\eta_1)}\\
a_2 &=& -\frac{\pi\mu b^2}{|z_{13}|^2(1-2\eta_1)^2 a_1},
\end{eqnarray}
where the special function $\gamma$ is a ratio of $\Gamma$ functions,
\begin{equation}
\gamma(x)=\frac{\Gamma(1-x)}{\Gamma(x)}.
\end{equation}
The conformal dimensions of the operators in the semiclassical limit are given by ($\alpha_i=\eta_i/b$)
\begin{equation}
h_i=\alpha_i(Q-\alpha_i) \approx \frac{\eta_i(1-\eta_i)}{b^2},
\end{equation}
and  the energy-momentum tensor can indeed be computed to be
\begin{eqnarray}
T(z) &=& \sum_{i=1}^3 \frac{h_i}{(z-z_i)^2}-[\frac{h_1+h_2-h_3}{(z-z_1)(z-z_2)}+\text{cyclic permutations}] \, .
 \label{classicalvevTO3}
\label{Tthree}
\end{eqnarray}
We verify this prediction of holomorphy explicitly in  appendix \ref{threeT} using the semi-classical solution (\ref{eq:classicalsolution}).

In summary, through the relation between the Einstein-Hilbert action in the bulk and the Liouville theory on the boundary \cite{Coussaert:1995zp}, reviewed and complemented in appendix \ref{action}, and the regularization of the Liouville action (\ref{regLiouvilleaction}) which we take to be valid for Liouville quantum gravity as well, we equate the semi-classical approximation to the quantum Liouville three-point function \cite{Zamolodchikov:1995aa} with the semi-classical approximation to the gravity action evaluated on the geometry (\ref{Banadosmetric}) determined by the energy-momentum tensor (\ref{Tthree}). In this manner, Liouville theory   propels itself as a proposal for a quantization of this classical theory of gravity.

We recall that this proposal  satisfies a  non-trivial consistency check, in that three-point functions of states with opposite momentum both associated to a given space-time energy-momentum tensor and metric are related through a phase factor (essentially given by the reflection amplitude).

\subsection{One Loop Correction}
\label{oneloop}

The semi-classical expansion of Liouville theory can be continued to higher orders. In particular, at one loop one can compute the exact quantum conformal dimension of the Liouville vertex operators \cite{Menotti:2004xz}. The details are as follows. By regularizing the one-loop determinant around the classical three-point configuration in an $SL(2,\mathbb{C})$ covariant manner, and by
analyzing
the dependence of the resulting one-loop determinant on the insertion points of the three Liouville  operators, one finds the $O(b^2)$ correction to the semi-classical conformal dimension
of the operator insertions and reproduces the exact conformal dimension \cite{Menotti:2004xz}:
\begin{equation}
h_i = \alpha_i (Q-\alpha_i) = \frac{1}{b^2} \eta_i (1-\eta_i) + \eta_i \, . \label{oneloopcorrection}
\end{equation}
It is proven explicitly  in \cite{Menotti:2004xz} that there is no further correction at two loop order, and, from our knowledge of Liouville conformal field theory, we know the property
holds to all orders. Thus, the conformal dimension is one loop exact.

The holographic interpretation of the result is that integrating over gravitons  gives a one loop correction to the mass, which is  one loop exact in Liouville quantum gravity. Note that here, we again identify the bulk gravitational measure with the Liouville measure.\footnote{From section \ref{Tmetrics} and appendix \ref{action} it is clear that this  entails a choice of how to integrate over metrics determined by energy-momentum tensor components.}

Finally, we note  the following phenomenon. When we compute the minimal (real) conformal dimension using the leading order formula, we find $1/(4 b^2)$ which equals $c/24$ in the semi-classical limit
$b \rightarrow 0$. When we use the exact formula (\ref{oneloopcorrection}), with the correction linear in the momentum $\eta_i$, then the exact minimal conformal dimension $(c-1)/24$ is again recuperated. Since the latter value for the minimal conformal dimension is the origin of the shift in the minimal mass of a black hole, we again recognize this as a one loop quantum gravity effect.\footnote{
	While formula (\ref{insertionmonodromy}) suggests a point particle spectrum of width $1/c$ for conformal dimensions between $(c-1)/24$ to $c/24$, the one loop correction due to integration over the Liouville, radial mode suggests the interpretation we provided here. This remains to be rigorously substantiated.	
}
	
	\section{Conclusions}
	\label{conclusions}
Liouville quantum gravity is a consistent and unitary theory of quantum gravity. It is a gravitational theory because it is the boundary incarnation of an Einstein-Hilbert action in three-dimensional anti-de Sitter
space. It is quantum mechanically consistent because Liouville conformal field theory is. The theory is a peculiar example of quantum gravity. The Euclidean three-dimensional anti-de Sitter space is a solution to the classical equations of motion, but it does not correspond to a normalizable quantum state. The spectrum of normalizable states  consists of primary states whose associated metric has
% tweak
hyperbolic (black hole) monodromies and whose descendants are BTZ geometries dressed with boundary gravitons. The black hole mass formula receives one loop quantum corrections. The entropy of the primary
% tweak
BTZ geometries is zero, while an arbitrary state with given high mass has a degeneracy determined by the effective central charge equal to one, associated to the boundary gravitons. There is a unique non-trivial three-point function between the states in the theory which is determined by the Liouville conformal field theory. This translates into a universal bulk gravitational interaction which is non-perturbatively well-defined.
	
Classically, we can interpret Liouville theory as an effective field theory. Like gravity, it comes about as a universal and crucial low-energy subsector  of certain theories of quantum gravity in three dimensions. However, it is hard to interpret the quantum Liouville theory as a universal effective field theory.  The quantum Liouville theory corrections will differ from quantum corrections in other microscopic theories.  Indeed, we defined our quantum gravitational measure as coinciding with the boundary Liouville measure. This fixes  the definition of the microscopic theory.
	
Our quantum theory of gravity is isolated in the sense that it is hard to add matter  to the theory consistently with unitarity, and for sure, the interpretation of the quantum theory of gravity, if it existed, would drastically change in the presence of non-topological degrees of freedom. The property that the theory is isolated in theory space is  reminiscent of other theories of quantum gravity, like string theory, in which adding even a single degree of freedom in space-time is an equally challenging enterprise. Our model also deviates from the standard view on quantum gravity, which requires the theory to account for the Bekenstein-Hawking entropy. Our proposal  evades this requirement since  the three-dimensional  theory of gravity is topological and therefore  lacks  local degrees of freedom.

In summary, we believe we defined an interesting theory of three-dimensional quantum gravity and encourage its further exploration. One direction of further research is to advance the Lorentzian interpretation of our  Liouville quantum gravity. It would also be very interesting to understand even better the bulk counterpart to the definition of the theory in terms of the Liouville measure. A relevant analysis is in \cite{Kim:2015qoa}, where the Liouville Hilbert space is obtained after quantization of the bulk degrees of freedom, and imposing a powerful projection condition. Understanding the relation between the projection procedure and the Liouville measure would be useful. Independently, the analysis of the bulk measure in \cite{Cotler:2018zff} may also  be useful in the comparison of the bulk and boundary  measures. It is  clear that the central idea of our paper can be transplanted to a host of other three-dimensional theories of gravity, with supersymmetry, higher spins, W-symmetry, diffferent boundary conditions, et cetera. We believe we opened up an avenue to enlarge our perspective on theories of quantum gravity by bringing a known speculation to its rigorous conclusion -- we hope this  turns out to be useful.

	\section*{Acknowledgments}
	It is a pleasure to thank our colleagues  for creating a stimulating research environment.  In particular, we thank C. Bachas, M. Paulos, L. Susskind and T.N. Tomaras for informative discussions. NT would like to thank the Laboratoire de Physique de l'Ecole Normale Sup\'erieure for hospitality.

	\appendix
	
	\section{Lorentzian Conformal Field Theory and Metrics}
	\label{LorentzianMetrics}
	In this appendix, we discuss Lorentzian  metrics that solve Einstein's equation with negative cosmological constant. We obtain the metrics through analytic
	continuation from the Euclidean metrics described in section \ref{metrics}. In particular, we again exploit the Banados form of the metric (\ref{Banadosmetric}), now in Lorentzian signature. We determine  Lorentzian energy-momentum tensor components through analytic continuation. Plugging those into the Lorentzian Banados metric, we obtain interesting solutions to Einstein's equations. We discuss properties of these solutions, but leave their full investigation for future work.

	\subsection{The Lorentzian Metrics}
The Lorentzian Banados solution reads:
	\begin{equation}
	ds^2 = l^2 d \rho^2 + 4 G l (-T (dw^+)^2 - \bar{T} (d {w^-})^2) + (l^2 e^{2 \rho}+ 16 G^2 T \bar{T} e^{-2 \rho}) dw^+  d w^-
	\end{equation}
	where $T$ and $\bar{T}$ indicate real and independent components of the Lorentzian energy-momentum tensor, and $w^\pm$ are real coordinates. The first example metric is when
	$T$ and $\bar{T}$ are constant. We then reproduce the Lorentzian BTZ black hole metric.	A second example is given by plugging in the Lorentzian version of the
	planar energy-momentum tensor:
	\begin{equation}
	T_{L}^{(0,\infty)} = \frac{h}{ (z^+)^2} \qquad \bar{T}_{L}^{(0,\infty)} = \frac{h}{(z^-)^2} \, .
	\end{equation}
	We obtain a horizon:
	\begin{equation}
e^{	4 \rho_{hor}} = (4 G /l)^2 h^2 (z^+ z^-)^{-2} \, .
	\end{equation}
	The horizon radius goes to infinity on the lines $z^+=0$ and $z^-=0$. Where the lines intersect, the singularity is the product of the individual quadratic singularities, and becomes quartic.
	The horizon therefore forms Lorentzian mountain ranges with extreme peaks where the ranges cross. The singularities are located where the horizon intersects the boundary. When we introduce the cut-off on the radius, a ribbon in
	the form of a cross is cut out from the horizon surface.

	Generically, it is subtle to analytically continue a complex valued energy-momentum tensor $T_{zz}$ to a real-valued energy-momentum tensor component $T_{++}$. However, we have readied the expressions for the energy-momentum tensor components in section \ref{metrics} in order to render this task eminently feasible.	
	Thus, the energy-momentum tensors $T^{(2)}_c$ and $T^{(3)}_c$ become real when we replace the coordinates $(w,\bar{w})$ by real coordinates $(w^+,w^-)$, and choose the source points $w_i^\pm$ to be real. This gives rise to energy-momentum tensors
	\begin{eqnarray}
	T_{c,L}^{(2)} (w^+) &=& \frac{h}{4} \frac{\sin^2 \frac{w_1^+-w_2^+}{2} }{ \sin^2 \frac{w^+ - w_1^+}{2} \sin^2 \frac{w^+-w_2^+}{2}}  +\frac{c}{24} \, , \nonumber \\
	T_{c,L}^{(3)} &=&  h_1 \frac{\sin \frac{w_2^+-w_1^+}{2}}{2 \sin \frac{w^+-w^+_1}{2} \sin \frac{w^+-w^+_2}{2}}   \frac{\sin \frac{w_3^+-w^+_1}{2}}{2 \sin \frac{w^+-w^+_1}{2} \sin \frac{w^+-w^+_3}{2}} + \mbox{cyclic} + \frac{c}{24} \, .
	\end{eqnarray}
We offer a few generic observations:	
\begin{itemize}
	\item When one light-cone coordinate coincides between two insertions, e.g. $w_1^+=w_2^+$, then the light-cone energy-momentum tensor component $T(w^+)$ does not
	depend on the conformal dimensions $h_{1}$ nor $h_2$ (but only on $h_3$), and only on the difference $w_2^+-w_3^+=w_1^+-w_3^+$, and in precisely the way of a two-point energy-momentum tensor expectation value.
	\item When three insertions have the same light-cone coordinate, the energy-momentum tensor component is constant. This follows from translational invariance.
\end{itemize}
These facts are a consequence of  the operator product expansion.

Naively, the metric associated to the energy-momentum tensor $T_{c,L}^{(3)}$ should correspond to three continuously interacting black hole
states, but in trying to substantiate this claim, we encountered difficulties.
To identify the precise physical interpretation of these interesting Lorentzian metrics requires considerably more work.

	\section{The Action}
	\label{action}
	
This appendix reviews the derivation of the Liouville action from the bulk Einstein-Hilbert action, via the Chern-Simons action in the bulk and chiral Wess-Zumino-Witten models on the boundary
\cite{Coussaert:1995zp}. A crucial extra step that we discuss is the gauging of an anti-diagonal symmetry that enforces the gluing of the left and right moving zero modes.
	
	\subsection{{From} Einstein-Hilbert  to  Wess-Zumino-Witten}
	The Einstein-Hilbert action in three dimensions with a negative cosmological constant $\Lambda=-l^{-2}$ can be written as the difference of two $sl(2,\mathbb{R})$ Chern-Simons actions \cite{Achucarro:1987vz,Witten:1988hc},
	\begin{equation}
	S_{EH}[A,\tilde{A}]=S_{CS}[A]-S_{CS}[\tilde{A}],
	\end{equation}
	where the connection one-forms $A$ and $\tilde{A}$ are related to the dreibein $e_\mu^a$ and the (dualized) spin connection $\omega_\mu^a$ throught the formulas $A_\mu^a=\omega_\mu^a+\frac{1}{l}e_\mu^a$ and $\tilde{A}_\mu^a= \omega_\mu^a-\frac{1}{l}e_\mu^a$.
	The Chern-Simons action can be written in polar $t,r,\theta$ components as
	\begin{eqnarray}
	S_{CS}[A] &=& \frac{l}{16\pi G_N}\int_M \tr(A\wedge dA+\frac{2}{3}A\wedge A\wedge A)\\
	&=& \frac{l}{16\pi G_N}\int_M dtdrd\theta \tr(A_\theta \dot{A}_r-A_r \dot{A}_\theta+2A_0F_{r\theta}) \, .
	\end{eqnarray}
	The Brown-Henneaux boundary conditions at large radius $r$ on the fields  $A$ and $\tilde{A}$ are \cite{Brown:1986nw}
	\begin{equation}
	A\sim\left(\begin{array}{ccc}
	\frac{dr}{2r} & O(\frac{1}{r})\\
	\frac{r}{l}dx^+ & -\frac{dr}{2r}
	\end{array}\right),
	\quad
	\tilde{A}\sim\left(\begin{array}{ccc}
	-\frac{dr}{2r} & \frac{r}{l}dx^-\\
	O(\frac{1}{r}) & \frac{dr}{2r}
	\end{array}\right),
	\label{eq:brownhennauxbc}
	\end{equation}
	where $x^\pm=t\pm\theta$.
This implies that $A_-=\tilde{A}_+=0$ on the boundary. If one computes the variation of the action $\delta S_{EH}$ when the equation of motion and this boundary condition are satisfied one obtains a surface term $\frac{l}{16\pi G_N}\delta\int_{\Sigma_2} dtd\theta \tr(A_\theta^2+\tilde{A}_\theta^2)$, where  $\Sigma_2$ is the asymptotic surface at $r\to\infty$. In order to make the action and the boundary condition compatible, we add a surface term to the Einstein-Hilbert action. The action  becomes
	\begin{equation}
	S[A,\tilde{A}]=S_{CS}[A]-\frac{l}{16\pi G_N}\int_{\Sigma_2}dtd\theta \tr(A_\theta^2)-S_{CS}[\tilde{A}]-\frac{l}{16\pi G_N}\int_{\Sigma_2} dtd\theta \tr(\tilde{A}_\theta^2) \, .\label{eq:improvedaction}
	\end{equation}
The field components	$A_0$ and $\tilde{A}_0$ are Lagrange multipliers that implement the constraints $F_{r\theta}=\tilde{F}_{r\theta}=0$. Solving these constraints one gets
	\begin{eqnarray}
	A_i &=& G_1^{-1}\partial_iG_1,\nonumber\\
	\tilde{A} &=& G_2^{-1}\partial_iG_2,
	\end{eqnarray}
	for $i=r,\theta$, where $G_{1,2}$ are elements of $SL(2,\mathbb{R})$.\footnote{This is the solution only when one does not consider  holonomies. For the case with non-zero holonomies, we refer to  the Appendix of \cite{Henneaux:1999ib}.} On the boundary, the fields $G_{1,2}$ asymptotically behave as
	\begin{equation}
	G_1\sim g_1(t,\theta)\left(\begin{array}{ccc}
	\sqrt{r} & 0 \\
	0 & \frac{1}{\sqrt{r}}
	\end{array}\right),
	\quad
	G_2\sim g_2(t,\theta)\left(\begin{array}{ccc}
	\frac{1}{\sqrt{r}} & 0 \\
	0 & \sqrt{r}
	\end{array}\right),
	\end{equation}
	such that the Brown-Henneaux boundary condition \eqref{eq:brownhennauxbc} is satisfied.
	Exploiting this parameterization, the action \eqref{eq:improvedaction} can be written as the difference of two chiral WZW actions \cite{Coussaert:1995zp}
	\begin{equation}
	S[A,\tilde{A}]=\frac{l}{16\pi G_N}(S^R_{WZW}[g_1]-S^L_{WZW}[g_2]) \, .
	\end{equation}
	Given the relation between the Einstein-Hilbert action and Chern-Simons theory on the one hand, and Chern-Simons theory and chiral Wess-Zumino-Witten models on the other hand, this is as expected.
	The two chiral WZW actions are given by
	\begin{eqnarray}
	S^R_{WZW} &=& \int_{\Sigma_2}dtd\theta \tr[g_1^{-1}\partial_tg_1g_1^{-1}\partial_\theta g_1-(g_1^{-1}\partial_\theta g_1)^2]+\Gamma[g_1]\nonumber\\
	S^L_{WZW} &=& \int_{\Sigma_2}dtd\theta \tr[g_2^{-1}\partial_tg_2g_2^{-1}\partial_\theta g_2+(g_2^{-1}\partial_\theta g_2)^2]+\Gamma[g_2],
	\end{eqnarray}
	where $\Gamma[g]$ is the three-dimensional part of the WZW action. Note that the chiral WZW actions are invariant under the transformations
	\begin{eqnarray}
	g_1 &\to& h_1(t)g_1f_1(x^+)\nonumber\\
	g_2 &\to& h_2(t)g_2f_2(x^-),\label{eq:sl2rsymmetry}
	\end{eqnarray}
	where $h_{1,2},f_{1,2}\in SL(2,\mathbb{R})$.
	\subsubsection*{Combining two chiral WZW to a non-chiral WZW by gauging a symmetry}
	\label{gauging}
	We gauge a subgroup of the above symmetry group.  This subgroup consists of transformations
	\begin{eqnarray}
	g_1 &\to& h(t)g_1\nonumber\\
	g_2 &\to& h(t)g_2,
	\end{eqnarray}
	and will guarantee that the left and right zero modes of $g_1$ and $g_2$ are glued together. Since these transformations are already symmetries of the action, gauging the symmetry comes down
	to restricting the integral over configuration space -- we only integrate over the zero modes a single time. This can be implemented by defining the fields
	\begin{equation}
	g=g_1^{-1}g_2,\quad u=-g_2^{-1}\partial_\theta g_1 g_1^{-1}g_2-g_2^{-1}\partial_\theta g_2,
	\end{equation}
	which are invariant under the gauged subgroup. In terms of these fields, we find that the action can be written as
	\begin{equation}
	S[g,u]=\frac{l}{16\pi G_N} [\int_{\Sigma_2} dt d \theta \tr(ug^{-1}\partial_tg-\frac{1}{2}u^2-\frac{1}{2}(g^{-1}\partial_\theta g)^2)-\Gamma(g)].
	\end{equation}
	The measures are related by
	\begin{equation}
	DgDu=\frac{1}{V}Dg_1Dg_2,
	\end{equation}
	where $V$ is volume of the set $\{g(t)\in SL(2,\mathbb{R})\}$.
	Performing the functional integral over the field $u$, we get the non-chiral WZW action
	\begin{equation}
	S[g]=\frac{l}{16\pi G_N} [\int_{\Sigma_2} dtd \theta \tr(2g^{-1}\partial_+gg^{-1}\partial_-g)-\Gamma(g)],
	\end{equation}
	where $\partial_\pm=\frac{1}{2}(\partial_t\pm\partial_\theta)$. Crucially, we have eliminated a zero mode.
	
	\subsection{{From} Wess-Zumino-Witten to Liouville}
	The next step amounts to Drinfeld-Sokolov or Hamiltonian reduction, induced by the boundary conditions. Firstly, we apply Gauss decomposition to write the remaining field $g$ as
	\begin{equation}
	g=\left(\begin{array}{ccc}
	1 & X \\
	0 & 1
	\end{array}\right)
	\left(\begin{array}{ccc}
	\exp(\frac{1}{2}\varphi) & 0 \\
	0 & \exp(-\frac{1}{2}\varphi)
	\end{array}\right)
	\left(\begin{array}{ccc}
	1 & 0 \\
	Y & 1
	\end{array}\right).
	\end{equation}
	The action will then become
	\begin{equation}
	S[g]=\frac{l}{8\pi G_N} \int dtd\theta [\frac{1}{2}\partial_+\varphi\partial_-\varphi+2\partial_-X\partial_+Y\exp(-\varphi)].
	\label{eq:boundaryaction}
	\end{equation}
Consider now the action as defined on a cylinder $[t_1,t_2]\times S^1$ of finite height \cite{Coussaert:1995zp}. The boundary conditions \eqref{eq:brownhennauxbc} will impose the following constraints:
	\begin{equation}
	\partial_-X\exp(-\varphi)=1,\quad \partial_+Y\exp(-\varphi)=-1.\label{eq:constraint}
	\end{equation}
	In order to make the variation of the action  zero when the equation of motion is satisfied, and in the presence of such constraint on the boundary, one  improves the action by adding a term. The action then reads
	\begin{equation}
	S_{\text{improved}}[g]=S[g]-\frac{l}{8\pi G_N}\oint d\theta (X\partial_+Y+Y\partial_-X)\exp(-\varphi)|^{t_2}_{t_1} \, .
	\end{equation}
	Plugging in the constraints, the improved action becomes a Liouville action
	\begin{equation}
	S[\varphi]=\frac{l}{8\pi G_N} \int dtd\theta [\frac{1}{2}\partial_+\varphi\partial_-\varphi+2\exp(\varphi)] \, .\label{eq:liouvilleaction}
	\end{equation}
Alternatively, one can determine the final action as follows.	
	One derives the equation of motion for the field $\varphi$ from the action \eqref{eq:boundaryaction}, and finds
	\begin{equation}
	\partial_+\partial_-\varphi=-2\partial_-X\partial_+Y\exp(-\varphi) \, .
	\end{equation}
	Eliminating the fields $X$ and $Y$ using the constraint \eqref{eq:constraint}, one obtains
	\begin{equation}
	\partial_+\partial_-\varphi=2\exp(\varphi) \, ,
	\end{equation}
	which is the same as the equation of motion one gets starting from \eqref{eq:liouvilleaction}. Therefore the boundary action is the Liouville action \eqref{eq:liouvilleaction}.

	\section{The Three-point Energy-momentum Tensor}
	\label{threeT}
In this appendix, we demonstate explicitly that the semiclassical limit of the energy-momentum tensor in the presence of three insertions is given by formula (\ref{Tthree}), as it must be.
	The energy-momentum tensor of Liouville theory in the semiclassical limit is \cite{Zamolodchikov:1995aa}
	\begin{eqnarray}
	T(z) &=& -(\partial\phi)^2+Q\partial^2\phi \nonumber\\
	&=& \frac{1}{2b^2}[-\frac{1}{2}(\partial\varphi)^2+\partial^2\varphi],
	\end{eqnarray}
	where we have used that $\varphi=2b\phi$ and that $Q=\frac{1}{b}+b\to\frac{1}{b}$ in the semiclassical limit $b \rightarrow 0$.
	Plugging in the classical solution \eqref{eq:classicalsolution}, the energy-momentum tensor can be written as
	\begin{equation}
	T(z)=-\frac{a_1\psi_1^{\prime\prime}(z)\psi_1(\bar{z})+a_2\psi_2(z)^{\prime\prime}\psi_2(\bar{z})}
	{b^2(a_1\psi_1(z)\psi_1(\bar{z})+a_2\psi_2(z)\psi_2(\bar{z}))}.
	\end{equation}
	We will  show that
	\begin{equation}
	\frac{\psi_1^{\prime\prime}}{\psi_1}=\frac{\psi_2^{\prime\prime}}{\psi_2},\label{eq:psi12}
	\end{equation}
	and therefore
	\begin{equation}
	T(z)=-\frac{\psi_1^{\prime\prime}}{b^2\psi_1}.
	\end{equation}
	Using the explicit form of the field $\psi_1$ given in \eqref{eq:psi1}, we find
	\begin{eqnarray}
	\frac{\psi_1^{\prime\prime}}{\psi_1} &=& \frac{\eta_1(\eta_1-1)}{(z-z_1)^2}+\frac{(1-\eta_1-\eta_3)(-\eta_1-\eta_3)}{(z-z_2)^2}
	+\frac{\eta_3(\eta_3-1)}{(z-z_3)^2} \nonumber \\
	& & +\frac{2\eta_1(1-\eta_1-\eta_3)}{(z-z_1)(z-z_2)}+\frac{2\eta_1\eta_3}{(z-z_1)(z-z_3)}
	+\frac{2(1-\eta_1-\eta_3)\eta_3}{(z-z_2)(z-z_3)} \nonumber \\
	& & +2[(\frac{\eta_1}{z-z_1}+\frac{1-\eta_1-\eta_3}{z-z_2}+\frac{\eta_3}{z-z_3})\frac{\partial x}{\partial z}+\frac{\partial^2 x}{\partial z^2}]\frac{_2F_1^\prime}{_2F_1} \nonumber\\
	& & +(\frac{\partial x}{\partial z})^2\frac{_2F_1^{\prime\prime}}{_2F_1},
	\end{eqnarray}
	where the symbol $_2F_1$ stands for the hypergeometric function $_2F_1$ with arguments $_2F_1(\eta_1+\eta_3-\eta_2,\eta_1+\eta_2+\eta_3-1,2\eta_1,x)$.
	Using the differential equation the hypergeometric function satisfies, we can express the second order derivative of the hypergeometric function $_2F_1$ in terms of its first order derivative and itself, which yields
	\begin{eqnarray}
	\frac{\psi_1^{\prime\prime}}{\psi_1} &=& \frac{\eta_1(\eta_1-1)}{(z-z_1)^2}+\frac{(1-\eta_1-\eta_3)(-\eta_1-\eta_3)}{(z-z_2)^2}
	+\frac{\eta_3(\eta_3-1)}{(z-z_3)^2} \nonumber \\
	& & +\frac{2\eta_1(1-\eta_1-\eta_3)}{(z-z_1)(z-z_2)}+\frac{2\eta_1\eta_3}{(z-z_1)(z-z_3)}
	+\frac{2(1-\eta_1-\eta_3)\eta_3}{(z-z_2)(z-z_3)} \nonumber \\
	& & +2(\frac{\eta_1}{z-z_1}+\frac{1-\eta_1-\eta_3}{z-z_2}+\frac{\eta_3}{z-z_3})\frac{\partial x}{\partial z}\frac{_2F_1^\prime}{_2F_1} \nonumber\\
	& & +[\frac{\partial^2 x}{\partial z^2}+(\frac{\partial x}{\partial z})^2(\frac{2(\eta_1+\eta_3)}{1-x}-\frac{2\eta_1}{x(1-x)})]\frac{_2F_1^\prime}{_2F_1} \nonumber\\
	& & +(\frac{\partial x}{\partial z})^2 \frac{(\eta_1+\eta_3-\eta_2)(\eta_1+\eta_2+\eta_3-1)}{x(1-x)}.\label{eq:ratio}
	\end{eqnarray}
	Applying the definition of the quantity $x$ given in \eqref{eq:x}, we obtain:
	\begin{eqnarray}
	\frac{\partial x}{\partial z} &=& \frac{z_{12}z_{32}}{(z-z_2)^2z_{31}},\\
	\frac{\partial^2 x}{\partial z^2} &=& -\frac{2z_{12}z_{32}}{(z-z_2)^3z_{31}},\\
	1-x &=& \frac{(z-z_3)z_{21}}{(z-z_2)z_31},\\
	x(1-x) &=& \frac{(z-z_1)(z-z_3)z_{21}z_{32}}{(z-z_2)^2z_{31}^2}.
	\end{eqnarray}
	We therefore find
	\begin{eqnarray}
	& & \frac{\partial^2 x}{\partial z^2}+(\frac{\partial x}{\partial z})^2(\frac{2(\eta_1+\eta_3)}{1-x}-\frac{2\eta_1}{x(1-x)})\nonumber\\
	&=& -2\frac{\partial x}{\partial z}[\frac{1}{z-z_2}+\frac{(\eta_1+\eta_3)z_{32}}{(z-z_2)(z-z_3)}-\frac{\eta_1z_{31}}{(z-z_1)(z-z_3)}]\nonumber\\
	&=& -2\frac{\partial x}{\partial z}(\frac{\eta_1}{z-z_1}+\frac{1-\eta_1-\eta_3}{z-z_2}+\frac{\eta_3}{z-z_3}),
	\end{eqnarray}
	and thus the total coefficient in front of $\frac{_2F_1^\prime}{_2F_1}$ on the right hand side of \eqref{eq:ratio} is zero. One can also simplify the remaining terms and get
	\begin{eqnarray}
	\frac{\psi_1^{\prime\prime}}{\psi_1} &=& \frac{\eta_1(\eta_1-1)}{(z-z_1)^2}
	+\frac{\eta_2(\eta_2-1)}{(z-z_2)^2}+\frac{\eta_3(\eta_3-1)}{(z-z_3)^2}\nonumber\\
	& & -[\frac{\eta_1(\eta_1-1)+\eta_2(\eta_2-1)-\eta_3(\eta_3-1)}{(z-z_1)(z-z_2)}+\text{cyclic permutations}] \, .
	\end{eqnarray}
	Mutadis mutandis, equation (\ref{eq:psi12}) follows. The final result for the energy-momentum tensor then agrees
	 with equation (\ref{Tthree}).

	\bibliographystyle{JHEP}

\begin{thebibliography}{99}
		
		%\cite{Weinberg:1976xy}
		\bibitem{Weinberg:1976xy}
		S.~Weinberg,
		``Critical Phenomena for Field Theorists,'' In Zichichi, Antonino, ``Understanding the Fundamental Constituents of Matter''. The Subnuclear Series 14, 1.
		doi:10.1007/978-1-4684-0931-4
		%%CITATION = doi:10.1007/978-1-4684-0931-4_1;%%
		%17 citations counted in INSPIRE as of 06 Mar 2019
		
		%\cite{Scherk:1974ca}
		\bibitem{Scherk:1974ca}
		J.~Scherk and J.~H.~Schwarz,
		``Dual Models for Nonhadrons,''
		Nucl.\ Phys.\ B {\bf 81} (1974) 118.
		doi:10.1016/0550-3213(74)90010-8
		%%CITATION = doi:10.1016/0550-3213(74)90010-8;%%
		%630 citations counted in INSPIRE as of 06 Mar 2019
		
		
		%\cite{Yoneya:1974jg}
		\bibitem{Yoneya:1974jg}
		T.~Yoneya,
		``Connection of Dual Models to Electrodynamics and Gravidynamics,''
		Prog.\ Theor.\ Phys.\  {\bf 51} (1974) 1907.
		doi:10.1143/PTP.51.1907
		%%CITATION = doi:10.1143/PTP.51.1907;%%
		%204 citations counted in INSPIRE as of 06 Mar 2019
		
		%\cite{Maldacena:1997re}
		\bibitem{Maldacena:1997re}
		J.~M.~Maldacena,
		``The Large N limit of superconformal field theories and supergravity,''
		Int.\ J.\ Theor.\ Phys.\  {\bf 38} (1999) 1113
		[Adv.\ Theor.\ Math.\ Phys.\  {\bf 2} (1998) 231]
		doi:10.1023/A:1026654312961, 10.4310/ATMP.1998.v2.n2.a1
		[hep-th/9711200].
		%%CITATION = doi:10.1023/A:1026654312961, 10.4310/ATMP.1998.v2.n2.a1;%%
		%14377 citations counted in INSPIRE as of 13 Feb 2019
		
\bibitem{Staruskiewicz}		

A.~Staruszkiewicz,
``Gravitation theory in three-dimensional space,"
Acta Phys.\ Polon. 24 (1963).
		
		
		%\cite{Deser:1983tn}
		\bibitem{Deser:1983tn}
		S.~Deser, R.~Jackiw and G.~'t Hooft,
		``Three-Dimensional Einstein Gravity: Dynamics of Flat Space,''
		Annals Phys.\  {\bf 152} (1984) 220.
		doi:10.1016/0003-4916(84)90085-X
		%%CITATION = doi:10.1016/0003-4916(84)90085-X;%%
		%951 citations counted in INSPIRE as of 11 Feb 2019
		
			%\cite{Banados:1992wn}\cite{Banados:1992gq}
		\bibitem{Banados:1992wn}
		M.~Banados, C.~Teitelboim and J.~Zanelli,
		``The Black hole in three-dimensional space-time,''
		Phys.\ Rev.\ Lett.\  {\bf 69} (1992) 1849
		doi:10.1103/PhysRevLett.69.1849
		[hep-th/9204099].
		%%CITATION = doi:10.1103/PhysRevLett.69.1849;%%
		%2407 citations counted in INSPIRE as of 23 Nov 2018
		
		%\cite{Banados:1992gq}
		\bibitem{Banados:1992gq}
		M.~Banados, M.~Henneaux, C.~Teitelboim and J.~Zanelli,
		``Geometry of the (2+1) black hole,''
		Phys.\ Rev.\ D {\bf 48} (1993) 1506
		Erratum: [Phys.\ Rev.\ D {\bf 88} (2013) 069902]
		doi:10.1103/PhysRevD.48.1506, 10.1103/PhysRevD.88.069902
		[gr-qc/9302012].
		%%CITATION = doi:10.1103/PhysRevD.48.1506, 10.1103/PhysRevD.88.069902;%%
		%1390 citations counted in INSPIRE as of 23 Nov 2018
		
		%\cite{Strominger:1996sh}
		\bibitem{Strominger:1996sh}
		A.~Strominger and C.~Vafa,
		``Microscopic origin of the Bekenstein-Hawking entropy,''
		Phys.\ Lett.\ B {\bf 379} (1996) 99
		doi:10.1016/0370-2693(96)00345-0
		[hep-th/9601029].
		%%CITATION = doi:10.1016/0370-2693(96)00345-0;%%
		%2482 citations counted in INSPIRE as of 29 Nov 2018
		
		
		%\cite{Brown:1986nw}\cite{Strominger:1997eq}
		\bibitem{Brown:1986nw}
		J.~D.~Brown and M.~Henneaux,
		``Central Charges in the Canonical Realization of Asymptotic Symmetries: An Example from Three-Dimensional Gravity,''
		Commun.\ Math.\ Phys.\  {\bf 104} (1986) 207.
		doi:10.1007/BF01211590
		%%CITATION = doi:10.1007/BF01211590;%%
		%1596 citations counted in INSPIRE as of 29 Nov 2018
		
		%\cite{Cardy:1986ie}
		\bibitem{Cardy:1986ie}
		J.~L.~Cardy,
		``Operator Content of Two-Dimensional Conformally Invariant Theories,''
		Nucl.\ Phys.\ B {\bf 270} (1986) 186.
		doi:10.1016/0550-3213(86)90552-3
		%%CITATION = doi:10.1016/0550-3213(86)90552-3;%%
		%1211 citations counted in INSPIRE as of 13 Feb 2019
		
		
		%\cite{Strominger:1997eq}
		\bibitem{Strominger:1997eq}
		A.~Strominger,
		``Black hole entropy from near horizon microstates,''
		JHEP {\bf 9802} (1998) 009
		doi:10.1088/1126-6708/1998/02/009
		[hep-th/9712251].
		%%CITATION = doi:10.1088/1126-6708/1998/02/009;%%
		%808 citations counted in INSPIRE as of 23 Nov 2018
		
	
	
	%\cite{Deser:1983nh}
	\bibitem{Deser:1983nh}
	S.~Deser and R.~Jackiw,
	``Three-Dimensional Cosmological Gravity: Dynamics of Constant Curvature,''
	Annals Phys.\  {\bf 153} (1984) 405.
	doi:10.1016/0003-4916(84)90025-3
	%%CITATION = doi:10.1016/0003-4916(84)90025-3;%%
	%373 citations counted in INSPIRE as of 11 Feb 2019
	
	
	%\cite{Martinec:1998wm}
	\bibitem{Martinec:1998wm}
	E.~J.~Martinec,
	``Conformal field theory, geometry, and entropy,''
	hep-th/9809021.
	%%CITATION = HEP-TH/9809021;%%
	%88 citations counted in INSPIRE as of 23 Nov 2018
	
	%\cite{Achucarro:1987vz}\cite{Witten:1988hc}
	\bibitem{Achucarro:1987vz}
	A.~Achucarro and P.~K.~Townsend,
	``A Chern-Simons Action for Three-Dimensional anti-De Sitter Supergravity Theories,''
	Phys.\ Lett.\ B {\bf 180} (1986) 89.
	doi:10.1016/0370-2693(86)90140-1
	%%CITATION = doi:10.1016/0370-2693(86)90140-1;%%
	%845 citations counted in INSPIRE as of 23 Nov 2018
	
	%\cite{Witten:1988hc}
	\bibitem{Witten:1988hc}
	E.~Witten,
	``(2+1)-Dimensional Gravity as an Exactly Soluble System,''
	Nucl.\ Phys.\ B {\bf 311} (1988) 46.
	doi:10.1016/0550-3213(88)90143-5
	%%CITATION = doi:10.1016/0550-3213(88)90143-5;%%
	%1843 citations counted in INSPIRE as of 23 Nov 2018
		
		%\cite{Witten:2007kt}
		\bibitem{Witten:2007kt}
		E.~Witten,
		``Three-Dimensional Gravity Revisited,''
		arXiv:0706.3359 [hep-th].
		%%CITATION = ARXIV:0706.3359;%%
		%386 citations counted in INSPIRE as of 29 Nov 2018
		
		%\cite{Gaberdiel:2007ve}
		\bibitem{Gaberdiel:2007ve}
		M.~R.~Gaberdiel,
		``Constraints on extremal self-dual CFTs,''
		JHEP {\bf 0711} (2007) 087
		doi:10.1088/1126-6708/2007/11/087
		[arXiv:0707.4073 [hep-th]].
		%%CITATION = doi:10.1088/1126-6708/2007/11/087;%%
		%57 citations counted in INSPIRE as of 29 Nov 2018
		
		%\cite{Maloney:2007ud}
		\bibitem{Maloney:2007ud}
		A.~Maloney and E.~Witten,
		``Quantum Gravity Partition Functions in Three Dimensions,''
		JHEP {\bf 1002} (2010) 029
		doi:10.1007/JHEP02(2010)029
		[arXiv:0712.0155 [hep-th]].
		%%CITATION = doi:10.1007/JHEP02(2010)029;%%
		%262 citations counted in INSPIRE as of 07 Mar 2019
		
		%\cite{Cotler:2018zff}
\bibitem{Cotler:2018zff}
  J.~Cotler and K.~Jensen,
  ``A theory of reparameterizations for AdS$_3$ gravity,''
  JHEP {\bf 1902} (2019) 079
  doi:10.1007/JHEP02(2019)079
  [arXiv:1808.03263 [hep-th]].
  %%CITATION = doi:10.1007/JHEP02(2019)079;%%
  %46 citations counted in INSPIRE as of 11 Dec 2019
		
			%\cite{Coussaert:1995zp}
			\bibitem{Coussaert:1995zp}
			O.~Coussaert, M.~Henneaux and P.~van Driel,
			``The Asymptotic dynamics of three-dimensional Einstein gravity with a negative cosmological constant,''
			Class.\ Quant.\ Grav.\  {\bf 12} (1995) 2961
			doi:10.1088/0264-9381/12/12/012
			[gr-qc/9506019].
			%%CITATION = doi:10.1088/0264-9381/12/12/012;%%
			%275 citations counted in INSPIRE as of 23 Nov 2018
		
	
	%\cite{Verlinde:1989ua}
	\bibitem{Verlinde:1989ua}
	H.~L.~Verlinde,
	``Conformal Field Theory, 2-$D$ Quantum Gravity and Quantization of Teichmuller Space,''
	Nucl.\ Phys.\ B {\bf 337} (1990) 652.
	doi:10.1016/0550-3213(90)90510-K
	%%CITATION = doi:10.1016/0550-3213(90)90510-K;%%
	%150 citations counted in INSPIRE as of 11 Feb 2019
		
		%\cite{Krasnov:2000ia}\cite{Krasnov:2002rn}
		\bibitem{Krasnov:2000ia}
		K.~Krasnov,
		``3-D gravity, point particles and Liouville theory,''
		Class.\ Quant.\ Grav.\  {\bf 18} (2001) 1291
		doi:10.1088/0264-9381/18/7/311
		[hep-th/0008253].
		%%CITATION = doi:10.1088/0264-9381/18/7/311;%%
		%29 citations counted in INSPIRE as of 29 Nov 2018
		
		%\cite{Krasnov:2002rn}
		\bibitem{Krasnov:2002rn}
		K.~Krasnov,
		``Lambda less than 0 quantum gravity in (2+1)-dimensions. 2. Black hole creation by point particles,''
		Class.\ Quant.\ Grav.\  {\bf 19} (2002) 3999
		doi:10.1088/0264-9381/19/15/309
		[hep-th/0202117].
		%%CITATION = doi:10.1088/0264-9381/19/15/309;%%
		%11 citations counted in INSPIRE as of 29 Nov 2018
		
		
			%\cite{Kim:2014bga}\cite{Kim:2015bba}\cite{Kim:2015qoa}
		\bibitem{Kim:2014bga}
		J.~Kim and M.~Porrati,
		``Long string dynamics in pure gravity on AdS$_{3}$,''
		J.\ Exp.\ Theor.\ Phys.\  {\bf 120} (2015) no.3,  477
		doi:10.1134/S1063776115030097
		[arXiv:1410.3424 [hep-th]].
		%%CITATION = doi:10.1134/S1063776115030097;%%
		%4 citations counted in INSPIRE as of 08 Mar 2019
		
	
		%\cite{Kim:2015bba}\cite{Kim:2015qoa}
		\bibitem{Kim:2015bba}
		J.~Kim and M.~Porrati,
		``More on long string dynamics in gravity on AdS$_3$ : Spinning strings and rotating BTZ black holes,''
		Phys.\ Rev.\ D {\bf 91} (2015) no.12,  124061
		doi:10.1103/PhysRevD.91.124061
		[arXiv:1503.06875 [hep-th]].
		%%CITATION = doi:10.1103/PhysRevD.91.124061;%%
		%5 citations counted in INSPIRE as of 08 Mar 2019
	
		
		%\cite{Kim:2015qoa}
		\bibitem{Kim:2015qoa}
		J.~Kim and M.~Porrati,
		``On a Canonical Quantization of 3D Anti de Sitter Pure Gravity,''
		JHEP {\bf 1510} (2015) 096
		doi:10.1007/JHEP10(2015)096
		[arXiv:1508.03638 [hep-th]].
		%%CITATION = doi:10.1007/JHEP10(2015)096;%%
		%11 citations counted in INSPIRE as of 08 Mar 2019
		%\cite{Porrati:2015eha}
		
		%\cite{Jackson:2014nla,Turiaci:2016cvo}
		\bibitem{Jackson:2014nla}
		S.~Jackson, L.~McGough and H.~Verlinde,
		``Conformal Bootstrap, Universality and Gravitational Scattering,''
		Nucl.\ Phys.\ B {\bf 901} (2015) 382
		doi:10.1016/j.nuclphysb.2015.10.013
		[arXiv:1412.5205 [hep-th]].
		%%CITATION = doi:10.1016/j.nuclphysb.2015.10.013;%%
		%54 citations counted in INSPIRE as of 26 Mar 2019
		
		%\cite{Turiaci:2016cvo}
		\bibitem{Turiaci:2016cvo}
		G.~Turiaci and H.~Verlinde,
		``On CFT and Quantum Chaos,''
		JHEP {\bf 1612} (2016) 110
		doi:10.1007/JHEP12(2016)110
		[arXiv:1603.03020 [hep-th]].
		%%CITATION = doi:10.1007/JHEP12(2016)110;%%
		%46 citations counted in INSPIRE as of 26 Mar 2019
		
			\bibitem{Germani:2013sra}
		C.~Germani,
		``On the many saddle points description of quantum black holes,''
		Phys.\ Lett.\ B {\bf 733} (2014) 93
		doi:10.1016/j.physletb.2014.04.030
		[arXiv:1307.6238 [hep-th]].
		%%CITATION = doi:10.1016/j.physletb.2014.04.030;%%
		%7 citations counted in INSPIRE as of 20 Mar 2019
		
		%\cite{Witten:1998qj}
		\bibitem{Witten:1998qj}
		E.~Witten,
		``Anti-de Sitter space and holography,''
		Adv.\ Theor.\ Math.\ Phys.\  {\bf 2} (1998) 253
		doi:10.4310/ATMP.1998.v2.n2.a2
		[hep-th/9802150].
		%%CITATION = doi:10.4310/ATMP.1998.v2.n2.a2;%%
		%9367 citations counted in INSPIRE as of 28 Feb 2019
		
		%\cite{Banados:1998gg}
		\bibitem{Banados:1998gg}
		M.~Banados,
		``Three-dimensional quantum geometry and black holes,''
		AIP Conf.\ Proc.\  {\bf 484} (1999) no.1,  147
		doi:10.1063/1.59661
		[hep-th/9901148].
		%%CITATION = doi:10.1063/1.59661;%%
		%218 citations counted in INSPIRE as of 13 Feb 2019
		
		
		%\cite{Seiberg:1990eb}\cite{Teschner:2001rv}
	\bibitem{Seiberg:1990eb}
	N.~Seiberg,
	``Notes on quantum Liouville theory and quantum gravity,''
	Prog.\ Theor.\ Phys.\ Suppl.\  {\bf 102} (1990) 319.
	doi:10.1143/PTPS.102.319
	%%CITATION = doi:10.1143/PTPS.102.319;%%
	%367 citations counted in INSPIRE as of 29 Nov 2018
	
	%\cite{Teschner:2001rv}
	\bibitem{Teschner:2001rv}
	J.~Teschner,
	``Liouville theory revisited,''
	Class.\ Quant.\ Grav.\  {\bf 18} (2001) R153
	doi:10.1088/0264-9381/18/23/201
	[hep-th/0104158].
	%%CITATION = doi:10.1088/0264-9381/18/23/201;%%
	%284 citations counted in INSPIRE as of 29 Nov 2018
		
		%\cite{Ponsot:2000mt,Teschner:2003at}
		\bibitem{Ponsot:2000mt}
		B.~Ponsot and J.~Teschner,
		``Clebsch-Gordan and Racah-Wigner coefficients for a continuous series of representations of U(q)(sl(2,R)),''
		Commun.\ Math.\ Phys.\  {\bf 224} (2001) 613
		doi:10.1007/PL00005590
		[math/0007097 [math-qa]].
		%%CITATION = doi:10.1007/PL00005590;%%
		%127 citations counted in INSPIRE as of 11 Feb 2019
		
		%\cite{Teschner:2003at}
		\bibitem{Teschner:2003at}
		J.~Teschner,
		``From Liouville theory to the quantum geometry of Riemann surfaces,''
		hep-th/0308031.
		%%CITATION = HEP-TH/0308031;%%
		%46 citations counted in INSPIRE as of 11 Feb 2019
		
	
		
		
	
	%\cite{Dorn:1994xn}\cite{Zamolodchikov:1995aa}
	\bibitem{Dorn:1994xn}
	H.~Dorn and H.~J.~Otto,
	``Two and three point functions in Liouville theory,''
	Nucl.\ Phys.\ B {\bf 429} (1994) 375
	doi:10.1016/0550-3213(94)00352-1
	[hep-th/9403141].
	%%CITATION = doi:10.1016/0550-3213(94)00352-1;%%
	%308 citations counted in INSPIRE as of 23 Nov 2018
	
%\cite{Zamolodchikov:1995aa}
\bibitem{Zamolodchikov:1995aa}
A.~B.~Zamolodchikov and A.~B.~Zamolodchikov,
``Structure constants and conformal bootstrap in Liouville field theory,''
Nucl.\ Phys.\ B {\bf 477} (1996) 577
doi:10.1016/0550-3213(96)00351-3
[hep-th/9506136].
%%CITATION = doi:10.1016/0550-3213(96)00351-3;%%
%551 citations counted in INSPIRE as of 17 Dec 2018
	
	
		%\cite{Teschner:1995yf}\cite{Collier:2017shs}
		\bibitem{Teschner:1995yf}
		J.~Teschner,
		``On the Liouville three point function,''
		Phys.\ Lett.\ B {\bf 363} (1995) 65
		doi:10.1016/0370-2693(95)01200-A
		[hep-th/9507109].
		%%CITATION = doi:10.1016/0370-2693(95)01200-A;%%
		%154 citations counted in INSPIRE as of 23 Nov 2018
		
		
		%\cite{Collier:2017shs}
		\bibitem{Collier:2017shs}
		S.~Collier, P.~Kravchuk, Y.~H.~Lin and X.~Yin,
		``Bootstrapping the Spectral Function: On the Uniqueness of Liouville and the Universality of BTZ,''
		JHEP {\bf 1809} (2018) 150
		doi:10.1007/JHEP09(2018)150
		[arXiv:1702.00423 [hep-th]].
		%%CITATION = doi:10.1007/JHEP09(2018)150;%%
		%25 citations counted in INSPIRE as of 23 Nov 2018
		
		%\cite{Witten:1998zw}
		\bibitem{Witten:1998zw}
		E.~Witten,
		``Anti-de Sitter space, thermal phase transition, and confinement in gauge theories,''
		Adv.\ Theor.\ Math.\ Phys.\  {\bf 2} (1998) 505
		doi:10.4310/ATMP.1998.v2.n3.a3
		[hep-th/9803131].
		%%CITATION = doi:10.4310/ATMP.1998.v2.n3.a3;%%
		%2798 citations counted in INSPIRE as of 29 Nov 2018
		
		
		%\cite{Hawking:1974sw}
		\bibitem{Hawking:1974sw}
		S.~W.~Hawking,
		``Particle Creation by Black Holes,''
		Commun.\ Math.\ Phys.\  {\bf 43} (1975) 199
		Erratum: [Commun.\ Math.\ Phys.\  {\bf 46} (1976) 206].
		doi:10.1007/BF02345020, 10.1007/BF01608497
		%%CITATION = doi:10.1007/BF02345020, 10.1007/BF01608497;%%
		%7521 citations counted in INSPIRE as of 12 Mar 2019
		
		%\cite{Carlip:1994gc}
		\bibitem{Carlip:1994gc}
		S.~Carlip and C.~Teitelboim,
		``Aspects of black hole quantum mechanics and thermodynamics in (2+1)-dimensions,''
		Phys.\ Rev.\ D {\bf 51} (1995) 622
		doi:10.1103/PhysRevD.51.622
		[gr-qc/9405070].
		%%CITATION = doi:10.1103/PhysRevD.51.622;%%
		%146 citations counted in INSPIRE as of 29 Nov 2018
		
		%\cite{Menotti:2004xz}
		\bibitem{Menotti:2004xz}
		P.~Menotti and G.~Vajente,
	``Semiclassical and quantum Liouville theory on the sphere,''
		Nucl.\ Phys.\ B {\bf 709} (2005) 465
		doi:10.1016/j.nuclphysb.2004.12.014
		[hep-th/0411003].
		%%CITATION = doi:10.1016/j.nuclphysb.2004.12.014;%%
		%16 citations counted in INSPIRE as of 17 Dec 2018
		
	
		
		
		%\cite{Henneaux:1999ib}
		\bibitem{Henneaux:1999ib}
		M.~Henneaux, L.~Maoz and A.~Schwimmer,
		``Asymptotic dynamics and asymptotic symmetries of three-dimensional extended AdS supergravity,''
		Annals Phys.\  {\bf 282} (2000) 31
		doi:10.1006/aphy.2000.5994
		[hep-th/9910013].
		%%CITATION = doi:10.1006/aphy.2000.5994;%%
		%109 citations counted in INSPIRE as of 19 Feb 2019
		
	\end{thebibliography}

\end{document}